# Machine learning-based method for linearization and error compensation of an absolute rotary encoder

**Lorenzo Iafolla[1], Massimiliano Filipozzi[1], Sara Freund[1], Azhar Zam[2], Georg Rauter[3] and Philippe Claude Cattin[1]**

[1] Center for medical Image Analysis & Navigation (CIAN), Department of Biomedical Engineering, University of Basel, Allschwil, Switzerland
[2] Biomedical Laser and Optics Group (BLOG), Department of Biomedical Engineering, University of Basel, Allschwil, Switzerland
[3] Bio-Inspired RObots for MEDicine-Lab (BIROMED), Department of Biomedical Engineering, University of Basel, Allschwil, Switzerland

E-mail: lorenzo.iafolla@unibas.ch; lorenzo.iafolla@outlook.it

**Abstract**

The main objective of this work is to develop a miniaturized, high accuracy, single-turn absolute, rotary encoder called ASTRAS360. Its measurement principle is based on capturing an image that uniquely identifies the rotation angle. To evaluate this angle, the image first has to be classified into its sector based on its color, and only then can the angle be regressed. Inspired by machine learning, we built a calibration setup, able to generate labeled training data automatically. We used these training data to test, characterize, and compare several machine learning algorithms for the classification and the regression. In an additional experiment, we also characterized the tolerance of our rotary encoder to eccentric mounting. Our findings demonstrate that various algorithms can perform these tasks with high accuracy and reliability; furthermore, providing extra-inputs (e.g. rotation direction) allows the machine learning algorithms to compensate for the mechanical imperfections of the rotary encoder.

Keywords: machine learning; rotary encoder; angular sensor; deep learning; shadow sensors

## 1 Introduction

The objective of our research[1] is to develop a high accuracy, miniaturized, single-turn absolute, rotary encoder. Rotary encoders (REs) are used in mechanical systems where rotational angles have to be measured and/or controlled, such as robotics systems. For example, it enables tracking the shape of a robotic hand by placing small REs at each joint of its fingers [1]. Consequently, REs are topic of research and new techniques are continuously proposed [2]-[5]. On the long term, our specific interest is in robotics for minimally-invasive surgery, where miniaturization is crucial. For many surgeries, e.g. laser osteotomy, measurement accuracy (better than 1 arcmin) is critical for achieving the desired treatment performance [6]-[8]. Also, absolute measurement is required for safety reasons because relative encoders would need recalibration by referencing, e.g. after a power failure. Although a wide variety of commercial absolute REs exists, their accuracy typically decreases with size, meaning that small absolute REs are not accurate enough for many applications. For example, among the best performing miniaturized absolute REs, there are the magnetic encoders, such as the AS5047U by AMS (8 mm diameter [9]) that can measure with an accuracy (limited by the systematic error) down to 2700 arcsec and precision (limited by the 14 bits resolution) of approx. 79 arcsec. Therefore, when absolute REs have to be used in small devices and high accuracy is required, there are little solutions available yet.

---
[1] This work was submitted for publication to "Measurement", Elsevier on the 7[th] July 2020.

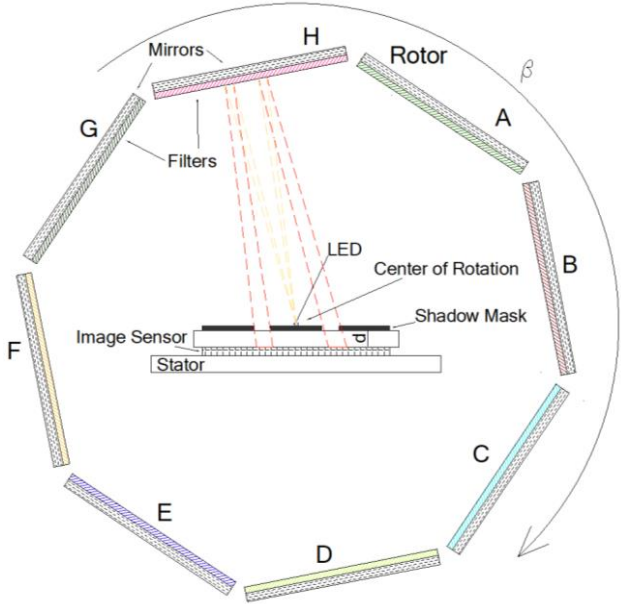

Figure 1. Section view of ASTRAS360. Each value of the rotation angle β corresponds to a unique shadow image captured by the image sensor.

In previous works, we presented a novel absolute angular measuring system (ASTRAS, Angular Sensor for TRAcking System, [10]) and its measuring interval (measurement range) extension to [0, 360] degrees (ASTRAS360, [11]). This type of measuring system belongs to the family of shadow sensors that measure the position of a light source by processing a shadow image cast onto an image sensor by a shadow mask [12]-[18]. ASTRAS consists of a rotor spinning about a stator. The stator hosts an image sensor, a shadow mask, and an LED. The rotor hosts a mirror that reflects the light from the LED, through the shadow mask, onto the image sensor. In ASTRAS360 (Figure 1), the rotor hosts several mirrors (eight in our case) in a way that the measuring interval is extended to 360 degrees. To distinguish the shadows associated with each mirror, color filters are placed in front of them to encode the light. Summarizing, the captured shadow image uniquely identifies the angle $β$ (measurand) between the rotor and the stator (see also Video 1 and 2 of the supplementary material). Compared to other absolute, optical, rotary encoders, ASTRAS360 features no fine grating track deposited on its rotor [19] and its high resolution relies only on the numerous, small-sized pixels of the image sensor.

In [10] and [11], measurement models of ASTRAS and ASTRAS360 were presented, providing an insight of the measuring systems, which is essential to effectively design the image processing algorithm and new implementations (e.g. the miniaturized version). In [20], a prototype of ASTRAS, featuring a miniaturized camera (NanEye by AMS, 1×1×0.5 mm$^3$, [21]) and a miniaturized shadow mask, was characterized. Its precision was 5 arcsec, corresponding to a 6σ-resolution (defined as six times the precision, [22]) equal to 6×5=30 arcsec (equivalent to 16 bits) which is a very promising result for future miniaturization of ASTRAS360.

However, some challenges related to image processing are still open. First, our measurement model is not accurate enough to define an accurate measurement function. This is particularly challenging for ASTRAS360 where the identification of the sector (i.e. the mirror in front of the stator, see Figure 2) is not straightforward due to optical artifacts unforeseen by the measurement model (e.g. the change of the color visible in Figure 3 and Video 1). Second, as demonstrated in [20], a miniaturized version of ASTRAS360 might not provide good quality images, prompting failures of the processing algorithm. Third, image processing must be performed in real-time for many applications (e.g. to close the feedback control loop of a robot), meaning that the algorithm should be, other than accurate, fast (few milliseconds). Fourth, the acquisition and processing system should be implemented preferably in a small embedded device or even on a single chip system (e.g. FPGA, [23]) avoiding bulky computational systems. Further-more, we are seeking for a processing algorithm able to compensate for the imperfections of the mechanics (e.g. play of the mechanical constraint) of our measuring system.

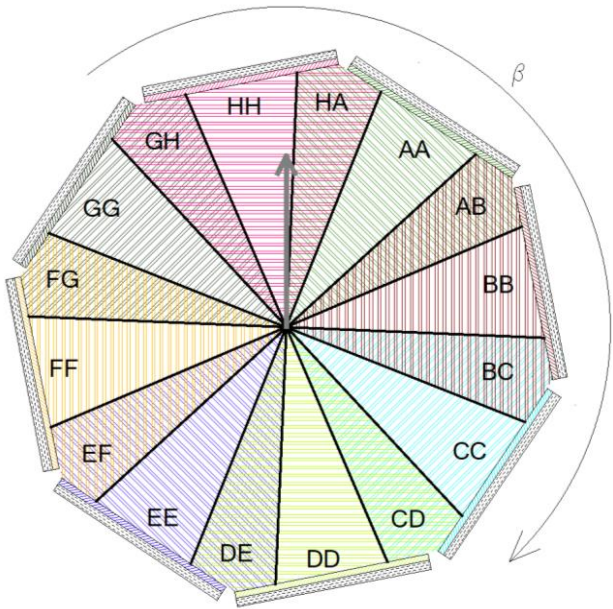

Figure 2. Section view of ASTRAS360 (stator is not represented, see Figure 1) showing the sectors. For the represented value of *β*, the sector would be HH as indicated by the grey arrow.

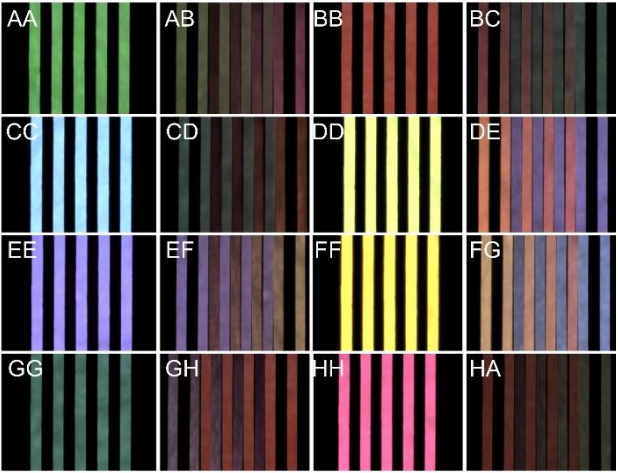

Figure 3. Examples of shadow images captured by ASTRAS360 corresponding to different sectors and different values of *β* (about 45 degrees apart). Should be noticed that the color of the light reflected by the same mirror changes with *β*; for example, the shadow in DD is yellow but the shadow originating from the same mirror is brownish in CD and DE.

Consequently, in this work, we wanted to study a set of solutions, based on Machine Learning (ML), to tackle the previous challenges. Compared to conventional (non-ML) measurement model algorithms (i.e. measurement functions), which are entirely defined and implemented by a human operator, ML algorithms make use of training data to automatically learn how to model the behavior of the system [24]-[26]. The advantage is that this model might be way more exhaustive, accurate, robust, and fast to compute than conventional approaches. For example, an ML algorithm can be trained to compensate for the imperfections of the mechanics of ASTRAS360. ML algorithms, and in particular Artificial Neural Networks (ANN), were already used for linearization and error compensation [27]-[33]. For instance, an ANN was used to compensate for the systematic error of a low-cost encoder in [27], and an ANN was used to compensate for the environmental influences (e.g. the temperature) of a pressure sensor in [28].

However, there is still a lot that ML algorithms cannot do by themselves. First, an ML algorithm must be selected according to the task and the requirements (accuracy, training time, inference time, memory, etc.). For example, the ML algorithms must be programmed to do a classification or a regression; in ASTRAS360, both types of tasks will be needed since the identification of the sector is a classification problem whereas the evaluation of *β* is a regression problem. Moreover, feature engineering must be done to achieve good results. In the ML jargon, a feature is one variable of the input (or a combination thereof); for example, if the input is an image, all its pixels are features. The input can also combine heterogeneous features like the pixels and the direction of rotation of ASTRAS360. The feature engineering consists in pre-processing the data in a way that the inputs will be workable and effective with the selected ML algorithm. Finally, a training dataset is needed to train the ML algorithm and, when the ML algorithm is supervised, also the desired outputs ("labels" in ML jargon) are needed.

To train and characterize the ML algorithms for ASTRAS360, we developed a calibration setup able to collect automatically several thousand measurements for different values of *β*. A commercial high-accuracy optical encoder (calibration standard) measured the value of *β* simultaneously, providing the desired output. For ML, this calibration setup is an automatic generator of labeled training data. To further test and characterize the ML algorithms and ASTRAS360, we collected data to assess the tolerance to the eccentricity of the rotor.

## 2 Material and methods

### 2.1 Measurement method of ASTRAS360

Figure 1 shows the schematic top view of ASTRAS360 and Figure 4, Videos 1, and 2 show photos, videos, and rendering of the prototype. A mechanical constraint, not shown in Figure 1, limits the degrees of freedom to rotation, identified by the angle *β*, of the rotor about the stator. The angle *β* is the measurand of ASTRAS360. The stator hosts an image sensor, a shadow mask, and an LED, whereas the rotor hosts several (eight in our case) mirrors with color filters in front. These components are arranged in a way that, one specific color and shape (i.e. a unique shadow image) corresponds to one *β* angle. This one-to-one correspondence is the feature that distinguishes an absolute encoder from a relative one. Examples of shadow images are visible in Video 1, which were recorded by the image sensor of ASTRAS360 while the rotor was rotating, i.e. *β* was changing.

In Figure 3, the labels AA, AB, BB, etc. denote the 16 sectors of ASTRAS360 (see also Figure 2). The shadow images indicated as AA, BB, CC, etc. are called single-shadow images and are generated by a single mirror. Those denoted as AB, BC, CD, etc. are called two-shadow images and are caused by two consecutive mirrors. The color of the shadow allows us to discriminate each sector (i.e. involved mirrors) from the others. However, notice that the color of the shadow from the same mirror changes with $\beta$ (see also Video 1).

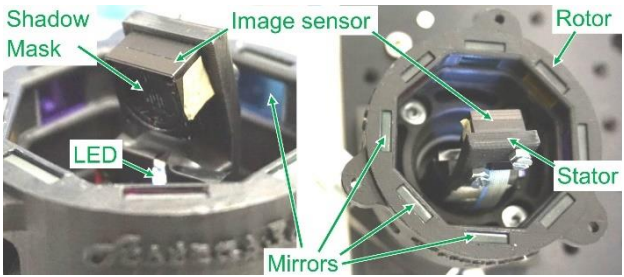

Figure 4. Side view (left) and top view (right) of the prototype of ASTRAS360. During the measurement operations a dome closes the system in order to block the external light.

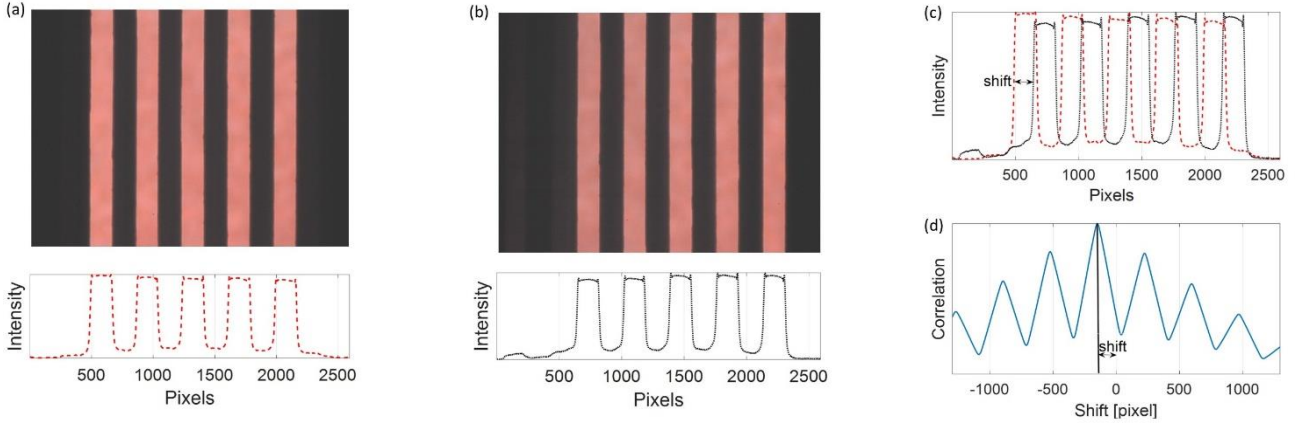

Figure 5. Shadow image (a) was the reference image of the sector BB ($d_{BB} \sim 12$ pixel/degree); below and in plot (c), the corresponding intensity vector (see Paragraph 2.3) is shown. Shadow image (b) was also captured in sector BB but 12.4 degrees apart from the reference; the corresponding intensity vector is plot below and in (c). The scatter plot (d) shows the correlation between the intensity vectors in (c); the position of its peak (~150 pixels) corresponds to $shift_{BB}$ in the measurement function (Equation (1)). We can calculate the position of the peak with sub-pixel resolution by fitting its neighborhood with a parabola [10]. Notice that the sign of the shift corresponds to the rotation direction of the rotor.

Whereas identifying the sector provides already a coarse measurement of $\beta$, its value is refined by comparing the input shadow image with a reference one captured during the calibration of ASTRAS360. Specifically, the only expected difference between the input and the reference image is a horizontal shift (see Video 1). This shift, measured in pixel units (i.e. the side length of a pixel), can be accurately determined by identifying the position of the peak of the correlation between the two images (Figure 5). In [11], it was demonstrated that the following measurement function, here defined for the sector AA, applies for all sectors:

$$\beta = arctan\left(\frac{shift_{AA}}{d_{AA}}\right) + \beta_{AA} \quad (1)$$

where $shift_{AA}$ is the shift defined above and in Figure 5, $\beta_{AA}$ is the reference angle at which the reference shadow image was captured for sector AA, and $d_{AA}$ defines the sensitivity of the system. It was also demonstrated in [10] and [11] that $d_{AA}$ is, in first approximation, the distance $d$ between the shadow mask and the image sensor (see Figure 1) although it might slightly change from sector to sector.

In summary, identifying the sector provides the corresponding values of $\beta_{AA}$ and $d_{AA}$ previously calculated from the calibration data, while $shift_{AA}$ can be measured with the correlation method. The angle $\beta$ can therefore be calculated using Equation (1) or, as it will be discussed in this paper, using better performing regression algorithms based on ML.

Finally, notice from Equation (1) and Figure 1 that theoretically, $\beta$ does not depend on small displacements of the center of rotation (eccentricity). However, we will show experimentally that this is only partially true in a way that ASTRAS360 can still work in presence of eccentricity, but its accuracy is affected.

### 2.2 Calibration setup

Figure 6 and Video 1 show the calibration setup. The prototype of ASTRAS360 (see also Figure 4 and Video 2) was mainly made of parts 3D-printed on a Fortus 250mc by Stratasys using an ABSplus-P430 thermoplastic. The radius of the rotor (distance between the mirrors and the center of rotation) was about 25 mm. The image sensor was a MU9PC by Ximea featuring 2592×1944 pixels with a size of 2.2×2.2 µm² [34]. The shadow mask was made of a layer of molybdenum 0.1 mm thick in which five 0.4 mm wide slits were laser cut. The distance between the image sensor and the shadow mask (*d* in Figure 1), which mainly defines the sensitivity, was about 1.1 mm. Further details are provided in [11].

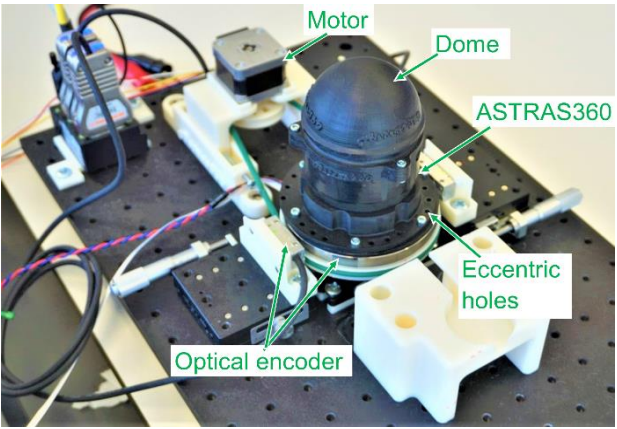

Figure 6. Calibration setup of ASTRAS360. During the measurements the measuring system was sealed with a dome in order to block the external light.

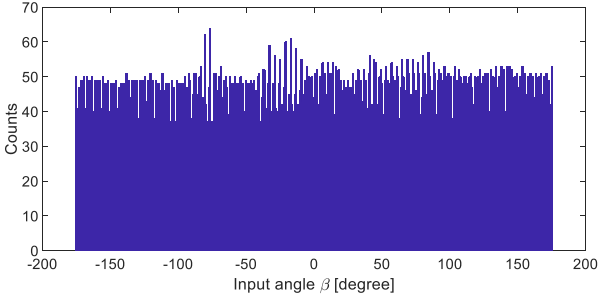

Figure 7. Distribution of the values of *β* of our dataset. Bin size is 1 degree.

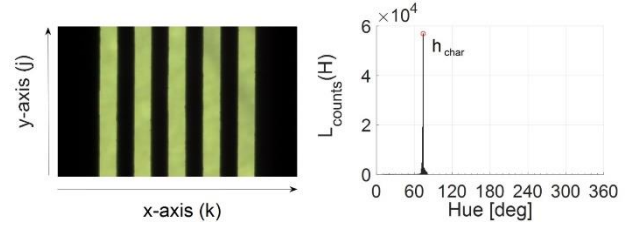

Figure 8. Shadow image (left) and corresponding weighted hue histogram (right).

In our setup, the rotor of ASTRAS360 was mounted on an interface plate with holes for fixation (eccentric holes in Figure 6 and Video 2). There are four sets of holes designed in a way that, with respect to those used during the calibration, we could change the eccentricity, i.e. displace the rotation center, by 1, 2, and 4 mm. The interface plate was fixed to the ring of a high accuracy, relative, optical, rotary encoder (Renishaw with Tonic T2011-15A read-head and REST20USA100 ring, [35]) which worked as calibration standard providing high accuracy (resolution 0.21 arcsec) measurements of *β*. These measurements are, in terms of ML, the desired outputs (or labels).

The rotor of ASTRAS360, the interface plate, and the ring of the optical rotary encoder were mounted over a rotary mount (Thorlab PR01/M, [36]), which acted as a mechanical constraint and limited the degrees of freedom to the rotation *β*. We coupled the rotary mount to a stepper-motor Sanyo Denki Sanmotion 103H5 to control the rotation (resolution ~0.8 degrees). We highlight here that the accuracy of the control was not crucial for our experiment; only the accurate measurement of the angle *β* was essential. We took advantage of the inaccuracy of the steps to collect samples at different *β*. To verify that the samples were approximately equally distributed over the full range of ASTRAS360, we used a histogram such as shown in Figure 7.

The rotary mount and all the parts above had a through-hole that allowed the stator (visible in Figure 4) to be fixed and to go through the optical rotary encoder and the rotor of ASTRAS360.

The acquisition of the signal from the optical rotary encoder was made by a microcontroller (Cortex-A9 MPCore processor embedded in a Xilinx Zynq Z7020) that recorded one sample per second and the corresponding timestamp. The same microcontroller was also driving the step-motor triggering a step every four seconds during the calibration. The type of optical rotary encoder we used could only measure angles between -180 and +180 degrees. To avoid output errors, we, therefore, programmed the microcontroller to limit the angles to -178 to 178 degrees. Consequently, we could not collect data for a range of about 4 degrees; however, we consider this acceptable as it represents only about 1% of the full angular range.

Finally, the data acquisition from the image sensor was done with a PC running software by Ximea (CamTool) that, during the calibration, could be programmed to acquire one image every four seconds and save it as a file named with a timestamp. Synchronizing the microcontroller and the PC with about 0.5 s accuracy, we were able to associate each image to the corresponding angle measured by the optical rotary encoder. The system could acquire autonomously several thousand samples working continuously for hours or even a few days.

### 2.3 Features engineering – data pre-processing

In this paragraph, we define the color intensity vectors ($I^R$, $I^G$, $I^B$), the histogram $L_{counts}(H)$, the intensity vector I, the mean intensity, and the direction of rotation which are the features used with the ML algorithms presented in Paragraph 2.4, 2.5, and 2.7.

The feature engineering was done based on the two following aspects: first, the sector identification requires features containing information related to the color, whereas the regression to calculate *β* does not; second, the rows of the shadow image can be averaged to improve the signal-to-noise ratio and to reduce the amount of data [11]. For this reason, rather than using the full images, we can use the color intensity vectors $I_k^R$, $I_k^G$, $I_k^B$ (or just $I^R$, $I^G$, $I^B$) where the superscript R, G, and B indicate the color channels (Red, Green and Blue) of the camera and *k* is an index which identifies the columns of the image (see Figure 8). For example, $I^R$ is calculated as following from the red channel of the shadow image, and similar equations also apply for the Green and Blue channels:

$$I_k^R = 100 \times \frac{\sum_{j=1}^{J} R_{jk}}{\max_{k'}\left(\sum_{j=1}^{J} R_{jk'}\right)} \quad (2)$$

where *j* identifies the rows of the image and $R_{jk}$ is the value of the pixel in *j*, *k*. J is the number of rows we want to average, in our case 1944.

As the sector identification relies on the color of the shadow image, it was useful to work with the HSI (Hue, Saturation,

Intensity, [37]) color representation instead of RGB. In particular, we used a weighted histogram $L_{counts}(H)$ (or just L) of the Hue defined as follows (see also Figure 8):

$$L_{counts}(H) = \sum_{\{k|H_k = H\}} (I_k^R + I_k^G + I_k^B) \quad (3)$$

where $H_k$ is the hue associated to $I_k^R, I_k^G, I_k^B$. $H$ is defined between 0 and 359, but its value can be discretized to obtain a histogram with a smaller number of bins.

The shift, defined in Paragraph 2.1, was the most effective feature for the regression algorithms. To measure it (see Figure 5), we used the intensity vector $I_k$ (or just I). This was obtained by averaging $I^R$, $I^G$, and $I^B$ as follow:

$$I = I_k = \frac{I_k^R + I_k^G + I_k^B}{3} \quad (4)$$

Notice that all intensity vectors are normalized to a maximum of 100. A further feature we used was the mean intensity of the image defined as $\Sigma_k(I_k)$.

Finally, we also used the direction of rotation (clockwise or counter-clockwise) of ASTRAS360 as a feature. This can be assessed from the previous value of the sector and of the shift. To transform the direction of rotation into a numerical number, as preferred by ML algorithms, we used the classic one-hot encoding [25]. This means that clockwise rotation was represented by ones and counter-clockwise by zeros.

## 2.4 Data labeling

The calibration setup described in Paragraph 2.2 could automatically associate each shadow image to the corresponding value of $\beta$ (label). However, as shown in Paragraph 2.1, the data processing also needed to classify the shadow images by sectors; therefore, we needed a further label that was not provided by the calibration setup.

One method, proposed in [11], consisted in distinguishing the two-shadow images from the single-shadow images. To do so, we first computed the intensity vector I (Equation (4)) and then we counted the number of its elements above a predefined threshold; the single-shadow images lay below the threshold and the two-shadow sectors above. Afterward, we associated each image to the corresponding sector according to their $\beta$.

The method we propose in this work is based on an unsupervised ML algorithm called k-Means [24]-[26] and we implemented it using the Matlab function kmeans. This algorithm tries to group the input data in k clusters by iteratively searching the centroid of each cluster. We used $\beta$ and the mean intensity of the shadow images (see Paragraph 2.3) as features. The parameter k was set to 16 because there were 16 sectors in our eight mirrors version of ASTRAS360 (see Figure 1 and Figure 3). We could also initialize the position of the centroids because we knew that the center of a single-shadow sector corresponded to the parallel position of the mirror with respect to the image sensor; instead, the center of a two-shadows sector was about 22.5 degrees apart from the centers of the contiguous single-shadow sectors.

It should be noticed that both these methods work only to label the calibration data because they provide the values of $\beta$, in other cases $\beta$ is unknown.

## 2.5 Sector classification methods

In Paragraph 2.1, we have shown that the first step of the data processing consists in identifying the sector; for example, an algorithm should be able to automatically associate each of the shadow images in Figure 3 to the corresponding sector. This is called classification problem.

In [11], a method based on analyzing the histogram L (see Equation (3) and Figure 8) was proposed. In particular, the position of its peak ($h_{char}$ in Figure 8), or of its two peaks for two-shadow images, was associated to a specific sector.

Seeking for more robust solutions, we investigated other methods based on ML and present here those that performed better which are: Convolutional Neural Network (CNN); k Nearest Neighbours (kNN); decision tree; and Support Vector Machine (SVM).

CNNs are feed-forward Convolutional Neural-Networks with hidden layers between the input and the output layers. A detailed description of the CNNs is provided in [24]-[26],[38],[39]. Shortly, the input of a CNNs is typically, but not necessarily, an image, and they can achieve astonishing results in classification; for example, they can classify objects, like cats or dogs, in an image. In our case, the inputs were the color intensity vectors, $I^R$, $I^G$, and $I^B$; these can be seen as the three channels of one RGB image with resolution equal to 1×K (K is the biggest value of $k$, see Paragraph 2.3, in our case it was 2592). We used the Deep Learning Toolbox of Matlab [39] to implement a CNN. Several hyper-parameters, such as the number of layers, have to be set depending on the task. Unfortunately, there is still not a systematic method to optimize the hyper-parameters; therefore, we just tried to simplify the CNN, e.g. reducing the number of layers, but still achieving the expected accuracy (100% for an absolute RE). In Table 1, we provide the hyper-parameters of our simplest, best performing, CNN.

kNN, decision trees, and SVM were not capable of classifying directly the color intensity vectors. More successful results were achieved using the histogram L (Equation (3)) as input. The number of bins of the histogram, that is the number of features, was minimized (see Paragraph 2.3) to improve the performances of the ML algorithm in terms of speed and memory, but still keeping the expected accuracy. To implement these ML algorithms, we used the Classification learner application of Matlab [40].

| Name of the layer | Parameters |
| --- | --- |
| imageInputLayer | [1 2592 3] |
| convolution2dLayer | [1 9], 16, 'Padding', 'same', 'Stride', [1 4] |
| batchNormalizationLayer | |
| reluLayer | |
| maxPooling2dLayer | [1 6], 'Stride', [1 9] |
| dropoutLayer | 0.2 |
| fullyConnectedLayer | 16 |
| softmaxLayer | |
| classificationLayer | |

Table 1. Architecture of the CNN used for the sector classification. For the name of the layers and the parameters, we refer to the documentation of Matlab [39].

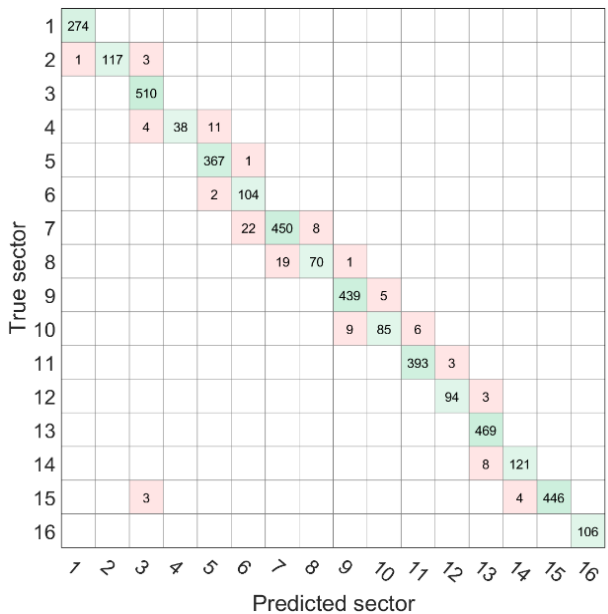

Figure 9. Confusion matrix obtained using kNN algorithm with k=100, holdout validation with 25% held out from 16786 samples. The false discoveries in (3,15) leads to a corrupted measurement of *β*. Despite this was not one of the best results, it corresponds to 99.9% accuracy.

kNN [26] is one of the oldest and simplest ML algorithm and is based on memorizing the training samples. Once a new sample is presented, it finds the k, which is a predefined integer, nearest neighbor in the space of the features among the training samples. The returning label is then decided by the majority of the k nearest neighbors. We performed tests with k equal to one, ten, and one hundred (see Figure 9 and Table 2).

A decision tree [26] is a sequential decision process that examines at each node a feature of the sample. Depending on its value, e.g. if it is above a predefined threshold, the algorithm will split and keep examining the sample following one branch after another. Once a leaf (i.e. the end of the branch) is reached, the decision about the class of the input is made. It should be noticed that the same feature can be examined several times. Therefore, the number of splits might be bigger than the number of features. In our case, we set the maximum number of splits to 100. Finally, a split criterion had to be chosen for the training; we used Gini's diversity index [40].

For a long time, SVM has been considered one of the best classifiers [26][40]. It tries to find the best separating hyperplanes in the space of the features: these hyperplanes are the decision boundaries allowing the classification of the samples.

It can also incorporate different kernels (linear, quadratic, cubic, Gaussian, etc.) to deal with non-linear data; however, in our case, a linear kernel was sufficient.

## 2.6 Comparing the classification methods

To assess the accuracy of the algorithms (ratio between correct predictions and the total number of samples, [24]), we analyzed the confusion matrixes of the testing subset, like that in Figure 9. In the confusion matrix, false predictions next to the diagonal are acceptable as they associate the shadow images to a sector contiguous to the true one. This does not affect the accuracy of the measurement as the corresponding Equation (1) still applies correctly to that shadow image. On the other side, when a shadow image is associated to a sector which is not contiguous to the true one, e.g. those three related to the cell (3,15) in Figure 9, the measurement is corrupted. Therefore, the accuracy has to be computed, accounting for only those three wrong predictions. For example, the accuracy related to the confusion matrix in Figure 9 is (4196-3)/4196=0.9993≡99.93%, where 4196 are the samples in the testing subset. The significance of this value should be evaluated considering that one single error leads to 99.98% accuracy. More advanced methods could be used to assess the accuracy of our ML algorithms with higher precision. For example, we could have used a n-folds validation method or have divided our data in training, validation, and testing subsets to avoid any bias [24]-[25]. However, for this work, we believe that our method is sufficient; indeed, more precise estimation of the accuracy would not add much value to the results because the accuracy is already very high.

In Paragraph 2.4, we introduced a method based on the unsupervised k-Means algorithm to label the data according to their sector. To assess its performance, we used a confusion matrix too, but in this case, the true sector was obtained with the non-ML method described in Paragraph 2.4.

It is not possible to assess the speed of prediction of an algorithm without having first selected the technology that will execute it. For example, when the ML algorithm can benefit from parallel computing, GPUs and FPGAs can perform better than CPUs. Furthermore, a real-time system should be used to have a deterministic assessment of the computational time. However, for a rough evaluation, we compare the performance by running the ML algorithms on a PC (Intel Xeon W-2133 CPU @ 3.60 GHz) with Matlab and forcing the execution only on the CPU and not on the GPU. Our interest focuses only on the prediction time as this will impact the real-time performances. Therefore, we do not discuss the training time. To assess the prediction time, we analyzed at least one hundred samples measured by the timer of Matlab (function tic).

Finally, to compare the required memory, we used the size of the model stored in Matlab workspace (function whos). This method, as well as that for the speed of prediction, is not the most accurate, however, it provides an initial indication of the differences between the algorithms. For example, it is easy to spot algorithms that memorize all the training data (e.g. kNN) and thus occupy more memory.

### 2.7 Regression methods

In this paragraph, we will present the regression ML algorithms that worked best for our problem; in particular, these were linear regression (polynomial fit), FFNN, and Gaussian processes.

It has been shown in Paragraph 2.1 and [11] that $\beta$ can be computed using Equation (1) once the sector has been identified (i.e. $\beta_{AA}$ and $d_{AA}$) and the $shift_{AA}$ has been measured (see Figure 5 and [10]). The parameters $\beta_{AA}$ and $d_{AA}$ can be evaluated from the mechanical specifications or from the calibration data using a model function regression such as the nlinfit function of Matlab. However, as Equation (1) is the outcome of a simple measurement model of ASTRAS360, it is not surprising that the accuracy is limited by a remaining systematic error (see Figure 12). For this reason, a method based on polynomial regression was proposed in [11]; in this case, $\beta$ can be computed, e.g. for sector AA, using the following:

$$\beta = \sum_{i=0}^{N} P_{i,AA} \cdot shift_{AA}^{i} + \beta_{AA} \qquad (5)$$

where $P_{i,AA}$ are the coefficients of the polynomial corresponding to sector AA. These coefficients have to be determined for each sector separately using the training data (calibration data) and can be done with the Matlab function polyfit. Polynomial regression is also an ML algorithm belonging to the family of the linear regression. Indeed, by adding some extra virtual features computed from the existing ones, e.g. their power, the linear regression can also fit data with strong non-linearity [26]. Therefore, in our case, the features are the shift and its powers up to the degree chosen according to the desired accuracy. To achieve the most accurate results, we used polynomials with a degree of up to eight for the two-shadow sectors and polynomials with a degree of up to twenty for the single-shadow sectors. This difference might be due to the different length of the sectors; two-shadow sectors span about 10 degrees and single-shadow span over 40 degrees.

We obtained accurate results also using an FFNN for regression (feedforwardnet function of Matlab). A detailed description of an FFNN is out of the scope of this paper, and the reader is referred to [24]-[26] and [38]. Nevertheless, it is important to know that such a network consists of layers with a certain number of nodes (neurons) that has to be decided by the operator during the calibration procedure. In our case, two-shadow sectors required up to nine neurons, and single-shadows sectors required up to eighteen neurons. In general, the number of layers can also vary, but for this specific task we used FFNN with only one hidden layer. The feature used as input was the shift. However, FFNN can be easily adapted to work with additional features; for example, we could improve the accuracy by using the direction of rotation (see Paragraph 2.3).

Gaussian processes [42] (fitrgp function in Matlab with exponential kernel and constant basis, [41]) was another ML algorithm that provided very accurate predictions using, as features, the shift and, optionally, the direction of rotation.

We also used regression methods to compute a coarse estimation of the shift (pre-shift) from the intensity vectors. Using the pre-shift, we could limit the calculation of the correlation in Figure 5, which is computationally very expensive, to a narrow neighborhood of the peak, improving the speed significantly. For example, we obtained reliable results with the general regression neural network method [43] implemented with the newgrnn function of Matlab (spread was set to one) and using I (Equation (4)) as input feature. As high accuracy is not required for this task, the pre-shift could be also calculated with several other methods not described in this paper.

Finally, it is worth mentioning how to prepare the training data. It is now clear that the dataset has to be split into subsets (one per sector); then, the ML algorithm has to be trained separately per each of them. To achieve the highest accuracy, in particular to reduce the peak-to-peak error, it was beneficial to add to each subset a few neighbor points (we added eight) from the adjacent sectors.

### 2.8 Comparing the regression methods

To compare the regression algorithms, we used the systematic error, which is the difference between the predicted angle and the input angle (measured by the optical rotary encoder).

We used scatter plots of the systematic error as a function of the input angle $\beta$ (e.g. Figure 10) to identify the sectors where the ML algorithm was not able to predict $\beta$ accurately. To highlight the errors due to variations over the time (e.g. caused by mechanical changes), we plotted the systematic error as a function of time (e.g. top of Figure 11). The regression fits well the data when the systematic error distributes randomly (e.g. Figure 12); otherwise, when it follows a specific pattern (e.g. curves like in Figure 10), it means that there is still space for improvement. To quantify the systematic error, we used its standard deviation and its peak-to-peak values.

Finally, we compared the algorithms in terms of speed and memory with the same approach as for the classification methods (Paragraph 2.6).

### 2.9 Eccentricity tolerance testing method

To test the eccentricity tolerance of ASTRAS360, we used the eccentric holes described in Paragraph 2.2. In this way, we could acquire four datasets (called Ecc0, Ecc1, Ecc2, and Ecc4) for four values of eccentricity (0, 1, 2, and 4 mm). We used the dataset Ecc0 for training and all the others for testing. In this way, we could simulate a misplacement of the rotation center with respect to its position during the calibration of ASTRAS360. For the sector classification, we used the CNN described in Table 1 and for the regression, we used the FFNN as described in Paragraph 2.7.

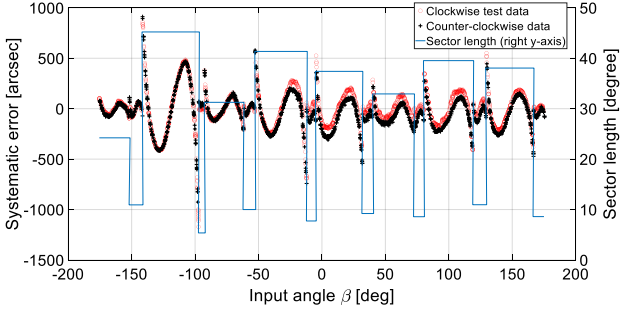

Figure 10. Systematic error associated to the model function of Equation (1). The continuous line (y-axis on the right) represents the sectors and their lengths. Training data are not displayed.

### 2.10  Training and testing datasets

To train and test the regression and classification algorithms, we used a dataset of 16786 samples (see the histogram in Figure 7). To avoid overfitting and wrong assessment of the accuracy, the dataset was divided into subsets as follows.

To train the classification algorithms (see Table 2), we divided the dataset into a training subset (75%, 12590 samples) and testing subset (25%, 4196 samples).

To train the regression algorithms (see Table 3), it was important to first identify the direction of rotation of the rotor. We called the samples collected with the rotor turning in clockwise direction "clockwise dataset" (they were 50% of the total), and the others "counter-clockwise dataset". Model function regression and polynomial regression algorithms were trained using a training subset (75%) of the clockwise dataset, whereas all the other data were used for testing. FFNN and exponential Gaussian processes were trained using 75% of the whole dataset.

Finally, to perform the eccentricity test, we trained the ML algorithms using a subset (9877 samples) of Ecc0, and we tested them using a held-out subset of Ecc0 (1090 samples) and all samples from Ecc1, Ecc2, and Ecc4 (see Table 4).

## 3  Results

### 3.1  Unsupervised labeling and classification results

In Paragraph 2.4, we introduced a method based on the k-Means algorithm to label the data according to their sector. As for the classification methods, we used a confusion matrix to assess its accuracy (Paragraph 2.6) that was 100%.

| ML algorithm name | Number of features | Time [ms] | Memory [kB] | Accuracy [%] |
|---|---|---|---|---|
| CNN 9 layers | 3×2592 | 2 | 100 | 100.0 |
| CNN 13 layers | 3×2592 | 2 | 60 | 100.0 |
| CNN 19 layers | 3×2592 | 2 | 450 | 100.0 |
| kNN (k=1) | 10 | 11 | 2100 | 100.0 |
| Decision tree | 23 | 6 | 50 | 100.0 |
| SVM | 15 | 35 | 760 | 100.0 |

Table 2. Performance of the classification algorithms.

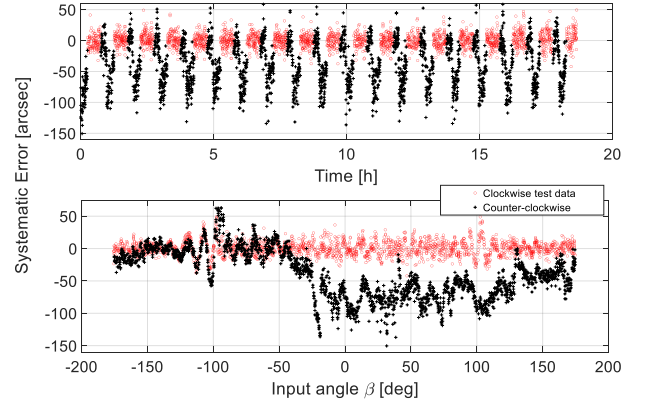

Figure 11. Systematic error associated to the polynomial regression method (Equation (5)). The training data are not displayed.

Table 2 provides the characterization results of the best performing classification algorithms. The number of features for the CNN algorithms was defined by the size of the color intensity vectors $I^R$, $I^G$, and $I^B$. The number of features for kNN, decision tree, and SVM corresponds to the number of bins of the histogram L. For these three algorithms, a higher number of features typically corresponded to higher accuracy; therefore, we reported the minimum number required to still achieve the best accuracy. Regardless of the number of bins, to compute the histogram L required 3.5 ms in addition to the prediction time of the ML algorithm under test. For example, the total time for the decision tree was 3.5+2.5 ms=6 ms.

### 3.2  Regression results

Table 3 provides the characterization results of the best performing regression algorithms. As an example of the methods to analyze the results introduced in Paragraph 2.8, we show the plots of the systematic errors in Figure 10 to Figure 12.

Figure 10 shows the systematic error associated to the model function regression (Equation (1)); in the same plot, we have shown, with a continuous blue line, how $\beta$ was divided in sectors. Specifically, the red circles are from the clockwise data set and the black crosses are from the counter-clockwise.

Figure 11 shows the systematic error versus the time (top) and input angle (bottom) associated to the polynomial algo-rithm (Equation (5)); in this case, we used eighth degree polynomial for two-shadow sectors and eighteenth degree polynomial for the single-shadow sectors. As before, the red circles are from the clockwise data set and the black crosses are from the counter-clockwise.

| Algorithm | Sigma [arcsec] | Peak-to-peak [arcsec] | Time [μs] | Memory [kB] |
|---|---|---|---|---|
| Model function regression | 187 | 2700 | <1 | 0.4 |
| Polynomial | 11 (157) | 136 (200) | <1 | 1.6 |
| Feed-forward neural-network | 8.6 | 96 | 60 | 670 |
| Exp. gaussian process | 8.3 | 267 | 30 | 1.2×10⁵ |

Table 3. Performance of the regression algorithms. For the method based on the polynomial fitting, we reported in parentheses the sigma and the peak-to-peak error associated to the counter-clockwise dataset.

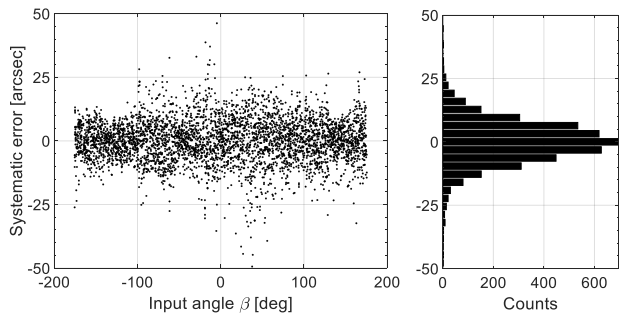

Figure 12. Systematic error and related histogram obtained using an FFNN. Clockwise and counter-clockwise data were combined; only the test data are shown in the plots.

Figure 12 shows the systematic error versus $\beta$ and its histogram obtained using an FFNN. In this case, and for the exponential Gaussian process, the features were the shift and the direction of rotation.

As a final remark, regression algorithms using features different from the shift (e.g. the intensity vectors) provided measurement less accurate or comparable to those of the model function of Equation (1); consequently the results are not reported here.

### 3.3 Eccentricity test results

The results of the eccentricity test, described in Paragraph 2.9, are reported in Table 4 and in Figure 13. The classification for Ecc0 dataset was performed with zero errors leading to 100% accuracy. For Ecc1 we got one error, for Ecc2 four errors, and for Ecc3 one error; these lead respectively to accuracy of 99.8%, 99.2%, and 99.8%. In the same table, we reported the standard deviation (sigma) and the peak-to-peak values of the systematic errors plotted in Figure 13.

## 4 Discussion

The sector classification method proposed in [11] was based on the analysis of the position $h_{char}$ of the peak (or the two peaks for the two shadow images) in the histogram L. It was successfully tested over a dataset of 444 samples, however, using the position of the peak deploys only a tiny portion of the information carried by the histogram L. Consequently, for some specific angular intervals, the results were not robust and a tedious fine tuning of the parameters of the algorithm was required. On the other side, the ML algorithms proposed in this work perform the classification using the whole histogram L (kNN, decision tree, and SVM) or directly the intensity vectors $I^R$, $I^G$, $I^B$ (CNN). Deploying all the information carried by these inputs, all algorithms were able to classify the sector with high accuracy and reliability. This was further demonstrated with the eccentricity test, where new images were correctly classified despite ASTRAS360 being assembled in slightly different configurations.

| Dataset | # of samples | Classification accuracy [%] | Sigma [arcsec] | Peak-to-peak [arcsec] |
|---|---|---|---|---|
| Ecc0 | 1090 | 100.0 | 10 | 106 |
| Ecc1 | 448 | 99.8 | 104 | 510 |
| Ecc2 | 446 | 99.2 | 165 | 970 |
| Ecc4 | 447 | 99.8 | 338 | 1800 |

Table 4. Performance of ASTRAS360 in presence of eccentricity. Sigma and peak-to-peak are referred to the systematic error shown in Figure 13. The precision of the classification accuracy is 0.2%.

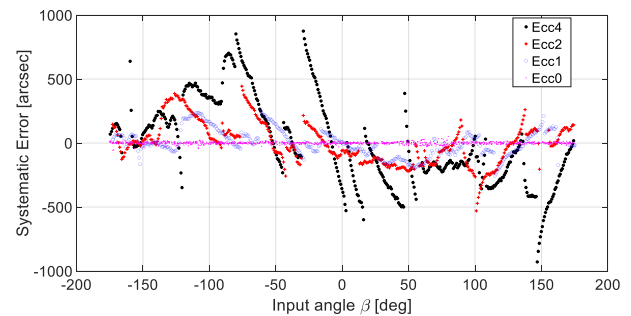

Figure 13. Systematic error with eccentricity equal to 0, 1, 2, and 4 mm. For Ecc0 we have plot only the test data.

Concerning the regression, we noticed that the measurements were affected by the rotation direction (see Figure 11). Most likely, this is due to play of the rotary mount axis caused by the changing forces applied by the transmission belt of the stepper-motor. We thus expect that the rotation axis of ASTRAS360 shifts slightly when the rotation direction changes or the platform hosting the rotor tilts; this affects differently the measurements of ASTRAS360 and of the calibration standard leading to a systematic error. This is a specific mechanical limitation of our 3D printed prototype and thus subject to change in severity depending on build quality. However, it is reasonable to foresee similar issues with new designs of ASTRAS360. In this regard, when the systematic error depends on a known variable, e.g. the rotation direction, we can use this variable as an additional input of the ML algorithm that will automatically compensate for the error. For example, in [28] and [30], the performance of a pressure sensor and a gyroscope could be improved using an FFNN with an extra input feature related to the temperature. In our case, we used the rotation direction as an extra input feature, enhancing the performance of our prototype of ASTRAS360. In general, future designs of ASTRAS360 might include additional sensors (e.g. thermometer or pressure sensor) to compensate for errors induced by environmental variables. From this point of view, we believe that the FFNN and Gaussian processes are superior to the polynomial fitting approach, as it is easier to embed an extra input feature. Indeed, using the same additional feature with the linear regression (which is the generalization of the polynomial method, see Paragraph 2.5), we did not achieve results comparable to those of the FFNN. One way could be to define one set of coefficients of the polynomial for each rotation direction by performing two separate calibrations; then, one set or the other would be used depending on the direction of rotation. However, this method would not be optimal for extra input features that change continuously such as the temperature.

Concerning the linearization of the output, the beneficial effect of using artificial neural networks has already been investigated in [27],[29],[31], and [32]. We concluded that the same should apply for ASTRAS360. However, we also showed that the polynomial method is one of the best performing in terms of accuracy, speed, and memory (see Table 3), although it is not straightforward to integrate extra input features. It is important to remark that using the shift as feature was the key to achieve accurate results in regression. Using different features led to poor results, demonstrating the importance of the feature engineering.

Finally, we have tested the tolerance of ASTRAS360 to eccentricity. This is an important issue for most of the high accuracy REs [44],[45]. Indeed, in the worst case, an optical encoder might not work even with eccentricities smaller than one millimeter. From data in Table 4, it is easy to calculate that the peak-to-peak systematic error of ASTRAS360 changes with the eccentricity at a rate of 425 arcsec/mm. In [44], it is shown that the peak-to-peak systematic error of an optical RE with the same radius (25 mm) changes in a range of 8250 arcsec/mm. To correct for this systematic error of optical REs [46] multiple read-heads are used; however, a similar technique could also be used for ASTRAS360 deploying multiple image sensors.

## 5   Conclusions

The main objective of our research is to develop a new, miniaturized, high accuracy, single-turn absolute, rotary encoder called ASTRAS360.

Pursuing this objective, in this work, we presented an experimental calibration setup for ASTRAS360. This setup was able to automatically collect thousands of samples alongside with rotation angle $\beta$ measured by a commercial RE. From the point of view of ML, we can see this system as an automatic generator of labelled training data. The data processing to calculate $\beta$ from the raw data included a classification problem and a regression problem. Using the collected dataset, we could demonstrate that several algorithms can reliably classify the data and accurately compute the angle $\beta$. The classification accuracy was 100%, and the standard deviation of the systematic error was smaller than 10 arcsec. We also presented the characterization of the best performing algorithms to provide a set of solutions for the future developments of ASTRAS360. For example, we want to develop a miniaturized version, with few millimeters diameter, using a miniaturized image sensor; we also want to develop a measuring system based on the same measurement principle but where the rotor can rotate about a pivot point (two degrees of freedom) rather than a single axis like in ASTRAS360 (one degree of freedom). In particular, this know-how will enable the design of new acquisition and processing systems able to fulfill the requirements of accuracy, reliability, speed, and usability in embedded systems. Finally, we demonstrated that ASTRAS360 is more tolerant to eccentricity than other optical REs.

With this work, we wanted to highlight the relevance of ML in measuring systems calibration and systematic error compensation. Considering miniaturization trend of microcontrollers, this method will become more and more relevant to the development of future accurate and smart measuring systems.

## Acknowledgments

This work was part of the MIRACLE project funded by the Werner Siemens-Foundation (Zug, Switzerland).